\documentclass[twocolumn,twoside,slac_two]{revtex4}
\usepackage{graphicx}
\usepackage{fancyhdr}
\begin{document}

\title{A long and homogeneous optical monitoring of the 'naked-eye' burst \\GRB 080319B with the Palomar-60 telescope}

%

\author{P. Veres}
\affiliation{E\"otv\"os University, Budapest, Bolyai Military University, Budapest}
\author{J. Kelemen}
\affiliation{Konkoly Observatory, Budapest}
\author{B. Cenko}
\affiliation{California Institute of Technology, Pasadena, University of California, Berkeley}
\author{Z. Bagoly}
\affiliation{E\"otv\"os University, Budapest}
\author{I. Horv\'ath}
\affiliation{Bolyai Military University, Budapest}
\begin{abstract}
	  GRB 080319B is one of the brightest and most extensively sampled bursts. It has good coverage at many wavelengths. Here we present the optical observations of the Palomar 60 inch telescope, which spans a long time interval after the burst. We augment the optical dataset with freely available Swift BAT and XRT observations reduced by us. We also compare our conclusions with the published parameters from the rich literature about this burst.
\end{abstract}

\maketitle
\thispagestyle{fancy}


\section{Introduction}
This is a Swift triggered burst and had a favorable position for early optical follow-up. Its redshift is z=0.937
P60 observations started at $T_0 + 170$ s and lasted for about $7$ days.
The main optical data set was observed with four filters with the Palomar - 60 inch telescope.
We revisit GRB080319B Palomar data, after it was published in part in \cite{2009ApJ...693.1484C}.
BAT and XRT data was obtained from the Swift websites. Reduction was carried out with the standard calibrating pipelines
batgrbproducts and xrtpipeline. The most recent calibration files were used. We also made use of the count data at
the XRT repository \citep{2007A&A...469..379E} and we used the count-to-flux-density conversion factor of
\cite{2007ApJ...663..407B}.

\section{Optical data reduction}
 The quick fading nature and in general the low average brightness of the
 GRB optical transients did not glow the automated (pipeline) reduction method,
 in the case of the "naked eye burst" oppositely the quite high brightness was
 the biggest problem. The shortcoming of automated methods  are the difficulty in:
 \begin{itemize}
 \item judging when to change from photometry of individual CCD frames
    to photometry of Co-added images
	\item  selecting the most reliable good seeing and good limiting magnitude
	   frames
	   \item selecting the best set of comparison stars both in alignment and
	      brightness range.
 \end{itemize}
		  Due to these problems and  the various alignment of the  neighbor objects,
		  the most reliable and the quickest way we found is the individual photometry
		  of the individual frames. In the present work therefore we choose the sturdy
		  aperture photometric method, allowing us to reach our goal in the quickest way.
		  When should we change from the photometry of the individual CCD frames to the
		  photometry of the co-added images.
		  The selection of the most reliable, good seeing, good limiting magnitude
		  frames is much easier in a manual way.
		   Selecting  the best set of comparison stars both in alignment and brightness
		   range is easier and quicker.\\
		   The non uniform nature of the optical CCD frames, twilight, dawn, Moon, bright
		   objects  nearby, focusing problems, filter inhomogeneities etc. are the
		   heaviest arguments beside our choice.\\
		      Therefore in view of the relatively small amount of observational data,
			  the advantage of the manual method over the pipeline method is clear. The
			  photometric accuracy of the obtained brightness data are in same range where
			  the accuracy as the other methods. In some cases where the OT faded near to
			  the observational threshold the errors are much higher due to the very low
			  signal to noise ratio. The only way to get much more precise data and longer
			  datasets is the usage of much bigger telescopes.

The slopes of the different observations show a gradual increase with decreasing energy band.

\begin{figure}[!t]
\centering
  \includegraphics[width=1.1\columnwidth]{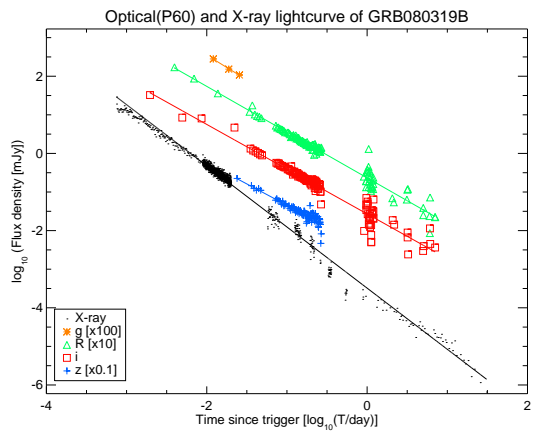}
  \caption{Joint $\gamma$-, X-ray and optical lightcurve of the naked-eye burst with simple power-law fits.}
\end{figure}

\begin{table}[t]
\begin{center}
\caption{}
\begin{tabular}{|c|c|c|}
\hline
 Band & Slope & Error\\ \hline
X-ray &        -1.58 & 0.01\\
 g-band&       -1.28 & 0.07 \\
 R-band&       -1.18&  0.02\\
 i-band&       -1.16& 0.03\\
 z-band&       -1.11& 0.05\\ \hline
\end{tabular}
\end{center}
\end{table}

\begin{figure}[!htb]
  \includegraphics[width=.99\columnwidth]{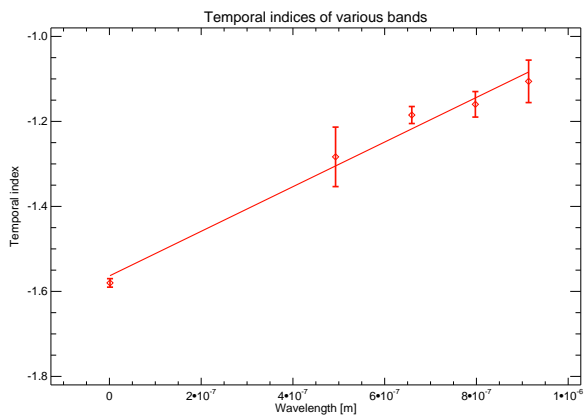}
  \caption{The spectral indices plotted against wavelength.}
\end{figure}

\section{Other measurements}
      \begin{figure}[htb]
          \centering
          \includegraphics[width=.99\columnwidth]{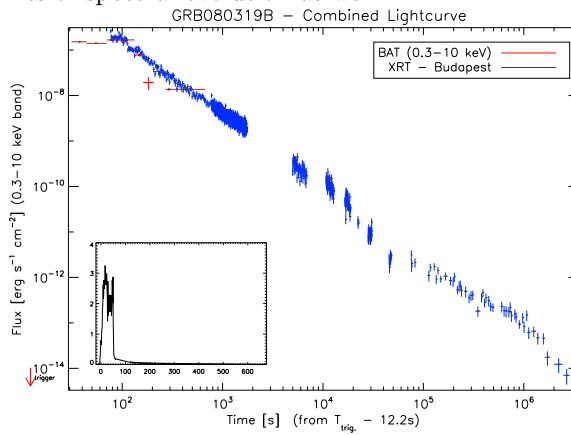}
          \caption{BAT and X-ray measurements in the $0.3-10$ keV range. The inset shows the BAT lightcurve  in the $15$-$150$ keV range.}
        \end{figure}
        There is still significant emission in the BAT band up to $660$ seconds after the trigger.
        The extrapolated BAT flux seems to match well the XRT flux. On closer inspection we find that BAT    photon indices are softer than the XRT indices in
        the five intervals of coincident measurement ($\Gamma_{\gamma} \simeq 2, \Gamma_{X}\simeq 1.7$).     There are small hints of spectral evolution as well.

\section{Discussion}
     GRB080319B is the brightest burst with known redshift. Without considering the two-component jet model  \citep{2008Natur.455..183R},
    we have fitted a power-law for the X-ray and the four optical bands. There is a hint of breaks only in   the XRT lightcurve, the optical bands seem to follow a
    power-law decline.
    We found that the temporal indices scale approximately linearly with wavelength.
    X-ray and the extrapolated $\gamma$-ray measurements seem to agree, but there is a hint a of a break     frequency.

\section{Acknowledgements}
        This research is supported by Hungarian OTKA grant K077795, by the Bolyai Scholarship (I. H.).

\bibliography{080319}


\end{document}